\documentclass[12pt,preprint]{aastex61}
\usepackage[utf8]{inputenc}
\usepackage{xcolor}
\usepackage{graphicx}
\usepackage{natbib}
\usepackage{amsmath}
\usepackage{lipsum}
\RequirePackage{lineno}
\renewcommand{\thefootnote}{\fnsymbol{footnote}}

\def\apjs{{ApJS}}
\def\apjl{{ApJL}}

\def\aap{{A\&A}}

\begin{document} 

\title{Updated Compositional Models of the TRAPPIST-1 Planets}
\author[0000-0001-8991-3110]{Cayman T. Unterborn}
\affiliation{School of Earth and Space Exploration, Arizona State University, Tempe, AZ, 85287, USA}

\author{Natalie R. Hinkel}
\affiliation{Department of Physics \& Astronomy, Vanderbilt University, Nashville, TN 37235, USA}

\author{Steven J. Desch}
\affiliation{School of Earth and Space Exploration, Arizona State University, Tempe, AZ, 85287, USA}

\correspondingauthor{Cayman Unterborn}
\email{cayman.unterborn@gmail.com}




\section{}
After publication of our initial mass-radius-composition (MRC) models \citep{Unte18} for the TRAPPIST-1 system, planet masses were updated in \citet{Grimm18}. We adopted the data set of \citet{Wang17} who reported different densities than \citet{Grimm18}. The differences in observed density change the inferred volatile content of the planets. \citet{Grimm18} report TRAPPIST-1b, d, f, g, and h as being consistent with $<$5 wt\% water and TRAPPIST-1c and e having largely rocky interiors. Here, we present updated results recalculating water fractions and potential alternative compositions using \citet{Grimm18} masses. 

\begin{figure}
    \centering
    \includegraphics[width=.55\linewidth]{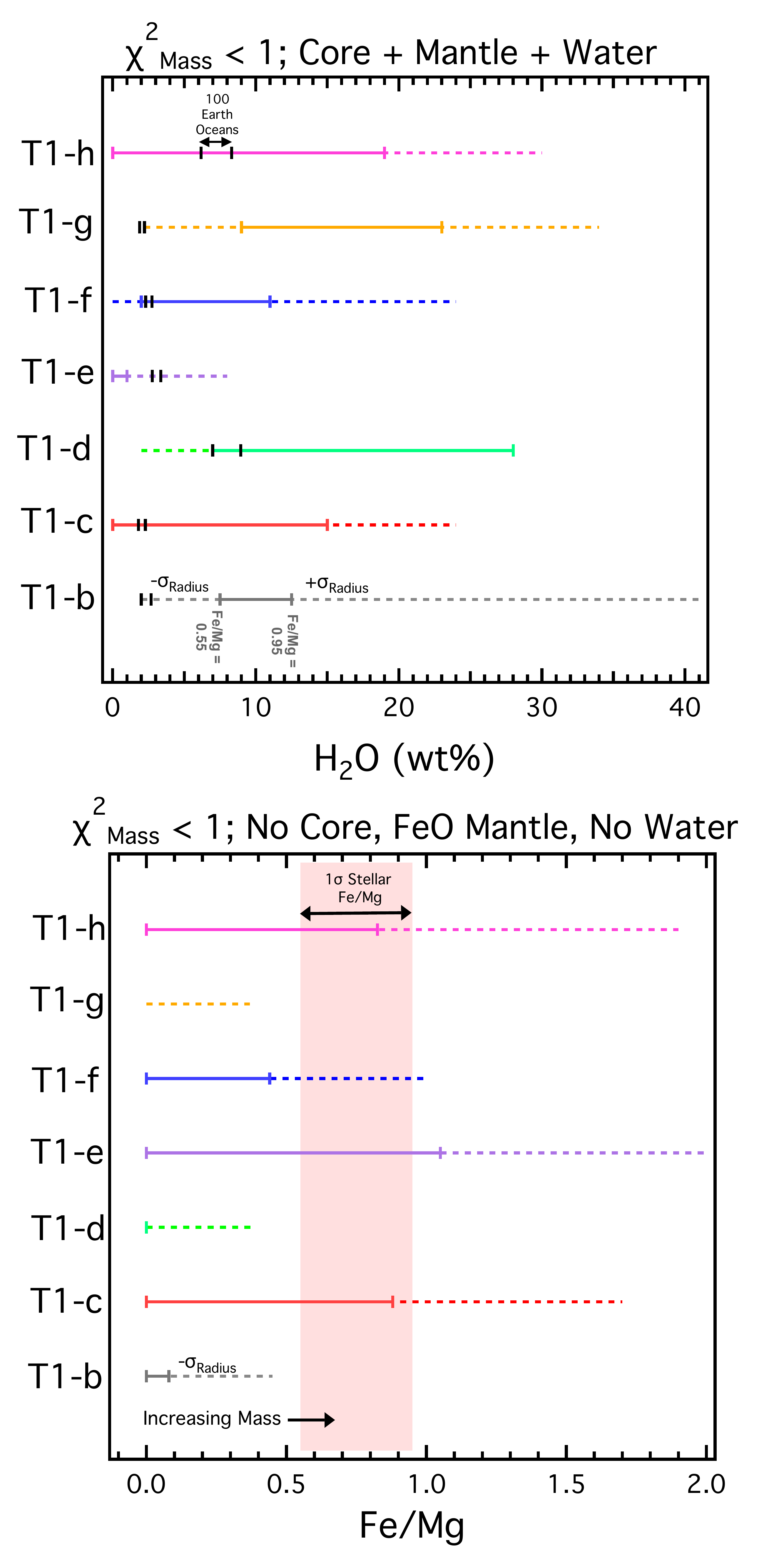}
    \caption{\textbf{Top:} Best fit water wt\% where $\chi_{\textrm{mass}}^2<1$ for each of the TRAPPIST-1 (T1) planets assuming a core, mantle, and water-ice layer. Solid and dashed lines represent our end-member core-to-mantle ratios (as Fe/Mg) including and not including any radius uncertainty, respectively. Black lines are the range of wt\% H$_2$O that produce 100 Earth ocean masses of water (1 Earth ocean = $1.2*10^{21}$ kg). \textbf{Bottom}: Best fit Fe/Mg for completely core-less planets using same format as above. The red band is the 1$\sigma$ range of stellar Fe/Mg for a sample similar metallicity stars as T1 \citep{Unte18}. If stellar composition broadly reflects planet composition, this represents a reasonable range of rocky planet Fe/Mg.}
    \label{Chi_squared}
\end{figure}

We find using the ExoPlex MRC code for a 3-layer planet (liquid core, silicate mantle with Mg/Si = 1, water-ice), planets b, d and g are consistent with the findings of \citet{Grimm18} of $<$5\% water by mass (Figure \ref{Chi_squared}, top), but only for the smallest core mass fraction planets in our models (23 wt\% Fe, Fe/Mg = 0.55). As the modeled core size increases, the relative amount of water needed to reduce the mass to within 1$\sigma$ of the observed mass ($\chi^2 < 1$) also increases. Planets c, e, f, and h are consistent with having no water, again only in the case that they have relatively small Fe-cores. When we adopt a larger core mass fraction (34 wt\%, Fe/Mg = 0.95) for planets c, e, f and h, we estimate these planets contain between 8--34 wt\% water.

To assign a probability of a planet having water, \citet{Grimm18} modeled a hypothetical composition where no Fe is present in the planet (Fe/Mg = 0). This composition, however, is unlikely as Fe, Mg, and Si have condensation temperatures very close to each other \citep{Unte17}, meaning if these planets contain silicates they likely contain iron as well. While stripping of a mantle is possible by giant impact after core segregation, in this case the stripped mantle would have to reform into a Fe-free planet. Therefore, Fe/Mg = 0 is not a realistic low-density end-member. 

One way for a planet with Fe/Mg above zero to have a lower bulk density without the addition of volatiles is for the planet's iron to alloy with ``light'' elements (e.g. Si, O) during core segregation. This process occurs under reducing conditions and is poorly understood for the Earth, whose core must contain 5--10 mol\% of these elements. Another mechanism for lowering planet density while retaining iron is to react the iron with oxygen prior to core formation and oxidize into FeO. This FeO will incorporate into Mg-bearing silicates and remain within the mantle. If all the iron oxidizes this way, the planet will have no core, but still contain significant Fe in the mantle. This is the true. Because this Fe-bearing silicates are more massive than the Mg-bearing alternative, the presence of Fe in the mantle will increase the density of the mantle. Therefore, the true lowest mass end-member composition for determining the likelihood of water is a fully oxidized planet with Fe/Mg$>$0.

We calculated MRC models for water-less, core-less compositions of the TRAPPIST-1 planets with Mg/Si = 1 (Figure \ref{Chi_squared}, bottom), varying relative mantle Fe up to Fe/Mg = 2. Planets b, d, and g can only be fit with compositions with Fe/Mg$<$0.5. These iron compositions fall below 1$\sigma$ of Fe/Mg abundance ratios from stars with similar metallicity to TRAPPIST-1 \citep[][Supplementary Figures 2 and 3]{Unte18}. Also as Fe/Mg increases, the modeled mass increases for the planets. From these points, we argue that planets b, d, and f \textit{must} have some significant volatile layer. The other planet's masses (c, e, f, and h) are well fit with any compositions  Fe/Mg$<$2, making the presence of water inconclusive. If these planets are indeed core-less, that is not to say they did not form wet because water would be the likely oxidant if it were present prior to core segregation. These planets may have indeed formed wet but chemical processes could lock this water within their mantles, producing no surface water.

Overall, we can reproduce the results of \citet{Grimm18} of planets b, d and g having $<5$ wt\% H$_2$O only if the cores of these planets are small ($<$23 wt\%). Due to the trade-off of increasing mass with increasing iron content at constant planet radius, if planets b, d, and f do have larger core mass fractions, more water must be present to lower their bulk density. Because these planets are still too massive, even in the core-less case, this strongly argues for the presence of planetary surface volatiles, likely water-ice. Planets c, e, f, and h, can either have volatile envelopes between 0--35 wt\% but are also consistent with being totally oxidized and core-less. While more refined mass and radius measurements of the TRAPPIST-1 planets will help further characterization, precise measurement of the host star composition will also aid in breaking this degeneracy in their interior compositions.

\bibliographystyle{aasjournal}
\bibliography{Trappist}

\end{document}